%% file: PersonaRelevanceJudgement.tex
\documentclass[sigconf]{acmart}

\usepackage{xcolor}
\usepackage{booktabs}
\usepackage{subcaption}
\usepackage{booktabs}
\usepackage{graphicx}   
\usepackage{xstring}    
\usepackage{booktabs}
\usepackage{multirow}
\usepackage{caption} 
\usepackage{tabularx}
\usepackage{placeins}

\newcommand{\mycaption}[1]{\caption{{\rm{#1}}}}

\definecolor{VibrantBlue}{rgb}{0.0, 0.2, 1.0}

\AtBeginDocument{%
  }

\setcopyright{acmlicensed}
\copyrightyear{2026}
\acmYear{2026}
\acmDOI{}
\acmConference[CIKM '26]{The 35th ACM International Conference on Information and Knowledge Management}{November 7--11, 2026}{Rome, Italy}

\usepackage{enumitem}
\usepackage{tcolorbox}

\begin{document}

\title{Persona Conditioning as an Assessor-Sensitivity Probe for LLM-Based IR Evaluation}

\author{Samaneh Mohtadi} 
\email{s.mohtadi@uq.edu.au}
\orcid{0009-0003-0980-6254}

\affiliation{%
  \institution{The University of Queensland}
  \city{Brisbane}
  \state{}
  \country{Australia}
}
\author{Pietro Bernardelle} 
\email{p.bernardelle@uq.edu.au}
\orcid{0009-0003-3657-9229}

\affiliation{%
  \institution{The University of Queensland}
  \city{Brisbane}
  \state{}
  \country{Australia}
}
\author{Joel Mackenzie} 
\email{joel.mackenzie@uq.edu.au}
\orcid{0000-0001-7992-4633}

\affiliation{%
  \institution{The University of Queensland}
  \city{Brisbane}
  \state{}
  \country{Australia}
}
\author{Gianluca Demartini} 
\email{g.demartini@uq.edu.au}
\orcid{0000-0002-7311-3693}
\affiliation{%
  \institution{The University of Queensland}
  \city{Brisbane}
  \state{}
  \country{Australia}
}

\begin{abstract}
Large language models (LLMs) are increasingly used as relevance assessors in information retrieval (IR) evaluation, raising questions about how assessor framing affects judgment reliability and downstream system comparison. We study persona conditioning as a diagnostic mechanism for exposing LLM assessor sensitivity. Using task-oriented personas drawn from two complementary sources (PersonaHub and NVIDIA Nemotron-Personas-USA), we instantiate five assessor roles emphasizing intent interpretation, domain expertise, contrastive judgment, evidence verification, and global search-quality assessment, compared with a standard UMBRELA baseline. Across six LLM backbones on TREC DL20 and RAG24, our analyses reveal structured rather than uniform assessor sensitivity. Judgments usually remain close to the baseline while shifting assessment strictness, evidential threshold, or interpretation emphasis rather than producing widespread relevance reversals. At the system level, high-capacity models preserve system-ranking agreement, while smaller models amplify persona-induced instability. Local rank-displacement analysis shows sensitivity concentrates on particular retrieval systems and system types, especially neural ranking/reranking systems on DL20 and RAG-oriented pipelines on RAG24. Persona source matters less than assessor role and model capacity. These findings position persona-conditioned judging as a controlled sensitivity probe for stress-testing LLM-based IR evaluation pipelines and identifying systems whose evaluation outcomes are sensitive to assessor framing.
\end{abstract}




\maketitle

\section{Introduction}
Relevance judgments play a central role in information retrieval (IR) evaluation by enabling system quality to be analyzed and compared. While traditionally produced by trained human assessors, the high cost and limited scalability of manual judging have driven growing interest in utilizing large language models (LLMs) as automated relevance judges. Recent work suggests that, under appropriate prompting, LLM judges can approximate professional assessors or real searcher preferences~\cite{thomas2024large,rahmani2025judging}. Other studies show that LLM-derived judgments can produce system rankings broadly comparable to those derived from human judgments, supporting scalable IR evaluation~\cite{upadhyay2024llmspatch,upadhyay2025umbrela}. However, LLM-based judgments are sensitive to prompt formulation and judging context~\cite{alaofi2024llms,clarke2025llm,soboroff2024dont}. They can also be affected by superficial document cues, thresholding choices, and evaluator bias~\cite{chen2024threshold,balog2025rankers,ye2024justice}. Recent guidelines further emphasize that these risks should be considered when using LLMs as judges~\cite{dietz2025principles}. These concerns motivate viewing LLM judges as components within a human--machine evaluation spectrum rather than as definitive sources of ground truth~\cite{faggioli2023perspectives}.

Beyond these concerns, many LLM-based judging approaches operationalize evaluation through a single fixed judging perspective. This perspective may approximate a professional assessor, a real searcher, or a standardized relevance guideline, but it still represents only one interpretation of the assessment task. In contrast, decades of IR research establish relevance assessment as inherently subjective~\cite{sormunen}, shaped by how assessors interpret query intent, task context, and relevance criteria~\cite{saracevic1996relevance,mizzaro1997relevance}. Large-scale evaluation studies show that assessor disagreement can affect evaluation outcomes, with structured variation in assessor preferences capable of meaningfully shifting system comparisons~\cite{voorhees2000variations,bailey2008relevance,carterette2010assessorerror}.
This observation motivates the following question for LLM-based IR evaluation: \textit{how sensitive are evaluation outcomes to changes in assessor perspective?}

Existing persona-conditioning work has explored psychological traits, demographic attributes, confidence styles, role-play prompts, and lexical prompt variation~\cite{jiang2023personality,chen2026personalityjudge,zheng2024persona}, while criteria-based approaches have proposed decomposing relevance into explicit dimensions such as accuracy, usefulness, and coverage~\cite{farzi2025criteria}. These efforts primarily target behavioral alignment or prompt optimization, leaving open how systematic variation in assessor perspective propagates through relevance labels, ranking stability, and retrieval-system evaluation outcomes.

In this paper, we study persona conditioning as an evaluator-sensitivity probe for LLM-based IR evaluation. Rather than treating personas as a mechanism for improving relevance labels, we use them to systematically perturb the assessor perspective and observe how these perturbations affect both relevance judgments and downstream IR system rankings. We instantiate task-oriented assessor roles covering query interpretation, domain expertise, contrastive judgment, evidence verification, and global search-quality assessment, and compare them against a standard UMBRELA judge~\cite{upadhyay2024umbrelasystem}. To examine whether persona source affects evaluation behavior, we compare abstract PersonaHub profiles~\cite{ge2024scaling} with skill-grounded personas from the NVIDIA Nemotron-Personas-USA collection~\cite{nvidia/Nemotron-Personas-USA}.

To enable scalable experimentation, we adopt summary-based judging and reuse fixed document summaries across persona runs, following prior work showing that concise summaries preserve judgment behavior and system-level stability while substantially reducing inference cost~\cite{mohtadi2026summarization}. We conduct experiments on the DL20~\cite{CraswellMMYC20} and RAG24~\cite{trec2024rag} datasets using multiple open and closed LLM backbones of different sizes. Our analysis examines persona effects at both the relevance-judgment and IR system ranking levels through graded agreement analysis, system ranking stability, and system-level rank-displacement analysis.

\medskip
Our study addresses the following research questions:
\begin{enumerate}[label={\textbf{RQ\arabic*}}, leftmargin=*]
\item\label{rq:label_effects}
How does persona conditioning affect relevance judgments relative to a standard UMBRELA-style LLM judge?
\item\label{rq:persona_source}
Do abstract personas and skill-grounded personas induce different evaluator-sensitivity patterns?
\item\label{rq:ranking_stability}
How stable are IR system rankings under persona-conditioned judging across datasets and model families?
\item\label{rq:system_impact}
Which retrieval systems are most influenced by persona-conditioned judging, and do these effects correlate with system characteristics?
\end{enumerate}
Persona conditioning produces structured, model-dependent sensitivity rather than uniform evaluation instability. At the judgment level, persona-conditioned labels generally remain close to the UMBRELA baseline, with differences appearing mainly as localized shifts in assessment strictness, evidential threshold, and interpretation emphasis. At the system level, sensitivity concentrates on particular retrieval-system types while global rankings remain relatively stable for high-capacity models. Persona source proves less consequential than assessor role and model capacity. These patterns position persona-conditioned judging as a controlled diagnostic stress test rather than a labeling alternative, with practical value for identifying retrieval systems whose evaluations are sensitive to assessor framing.
\section{Background and Related Work}
\paragraph{Relevance Subjectivity and Evaluation Stability.}
Relevance in IR is inherently subjective and shaped by user intent, task context, and assessor interpretation of relevance criteria~\cite{mizzaro1997relevance,saracevic1996relevance,sormunen}.
Empirical TREC analyses have repeatedly documented this variability. Voorhees~\cite{voorhees2000variations} showed that although individual assessors often disagree on relevance, evaluation outcomes can remain relatively stable when using different sets of qualified assessors. Zobel~\cite{zobel1998reliable} similarly examined the reliability of large-scale pooled evaluation and found that system comparisons are often robust to variation in relevance judgments. Subsequent work has refined this picture: Bailey et al.~\cite{bailey2008relevance} demonstrated that assessors of different quality levels do not produce interchangeable judgments and that assessor type affects measured effectiveness, while Carterette and Soboroff~\cite{carterette2010assessorerror} showed that systematic assessor error can shift system rankings even when label-level agreement remains high. Together, these findings motivate treating assessor variation not as annotation noise to be averaged out, but as a structured component of IR evaluation whose effects on downstream system comparison must be characterized.

\paragraph{LLMs as Relevance Judges.}
The use of LLMs as automated evaluators was first widely formalized in general NLP, where frameworks such as MT-Bench~\cite{zheng2023judging} and G-Eval~\cite{liu2023geval} showed that LLMs can produce evaluation scores correlating strongly with human judgments across open-ended tasks. Adapting these ideas to IR, the cost and scalability limitations of manual judging have motivated widespread exploration of LLMs as automated relevance assessors. Under carefully designed prompting protocols, LLM judges can approximate professional assessors or real searcher preferences~\cite{thomas2024large} and produce system rankings broadly comparable to those derived from human judgments~\cite{upadhyay2024llmspatch,upadhyay2025umbrela}. Frameworks such as LLMJudge formalize LLM-based relevance assessment pipelines and support systematic comparison of judging configurations~\cite{rahmani2025judging}, while standardized prompting approaches such as UMBRELA~\cite{upadhyay2024umbrelasystem,upadhyay2025umbrela} enable large-scale analysis of agreement patterns and evaluation behavior across datasets and models. 
However, the generalizability of these frameworks depends on the backbone model: Farzi and Dietz~\cite{farzi2025umbrela} reproduce UMBRELA across multiple LLMs and find that smaller models exhibit meaningfully degraded assessment performance relative to larger ones. Criteria-based approaches~\cite{farzi2025criteria} further decompose relevance into explicit assessment dimensions, such as accuracy, usefulness, and coverage. Recent work shows that input-side formulation influences evaluation behavior, as formalized topic descriptions and narratives substantially improve agreement with human judgments~\cite{keller2026formalized}. Benchmarking studies show that LLM-based relevance assessment is sensitive to experimental design choices, with outcomes varying across prompting strategies and judging setups~\cite{arabzadeh2025benchmarking}. Moreover, label-level agreement alone does not guarantee ranking stability, since similar agreement levels can still produce meaningfully different system orderings~\cite{gera2024justrank,bailey2008relevance,carterette2010assessorerror}. Our work differs from these criteria-based, input-side, and prompting-focused approaches by varying \emph{who} is assumed to be judging rather than \emph{what} is assessed or \emph{how} the assessment input is structured, while keeping the underlying judging framework fixed.

\paragraph{Bias and Instability in LLM-based Evaluation.}
LLM-based relevance judgment raises concerns about reliability, robustness, and bias, particularly when LLM-generated labels are used as replacements rather than supplements for human assessors~\cite{clarke2025llm,soboroff2024dont}. Broader NLP studies have documented systematic biases in LLM evaluators, most notably positional or order bias, in which evaluation outcomes can shift simply by changing the order in which candidate responses appear~\cite{wang2024fairevaluators}. Empirical studies show that LLM judgments are sensitive to prompt formulation, superficial document cues~\cite{yu26arxiv}, and judging context; for example, injecting query terms into irrelevant documents can inflate relevance labels without improving substantive relevance~\cite{alaofi2024llms,alaofi2026llmrelevance}. Cognitive-bias-like effects have also been identified in LLM assessors, including threshold priming~\cite{chen2024threshold} and recency bias~\cite{fang2025recency}. At the system level, LLM judges may favor LLM-based ranking and generation systems, raising concerns about circular evaluation and self-preference~\cite{balog2025rankers}. Broader analyses identify structural evaluation risks such as circularity, LLM narcissism, loss of variety of opinion, and multiple systematic judge biases~\cite{dietz2025principles,ye2024justice}. Recent work has begun to show that disagreement and evaluator behavior can exhibit structured patterns within the query--document representation space, enabling analysis beyond aggregate label-level agreement~\cite{mohtadi2026densevectors}. These studies typically characterize biases that arise within a fixed judging configuration; we instead examine sensitivity that emerges when the judging perspective itself is systematically varied.

\paragraph{Persona Conditioning and Assessor Perspective.}
Most directly relevant to our study, recent work examines how persona conditioning alters LLM behavior in evaluation tasks. Wang et al.~\cite{wang2026roleplay} show that role-play signals can systematically alter zero-shot ranking behavior while remaining only weakly entangled with query--document representations, suggesting that persona effects may driven by evaluator framing rather than changes to the underlying query--document representation. Work on synthetic annotators and crowd impersonation finds that persona conditioning does not reliably reproduce the diversity or inconsistency patterns observed in real human assessors~\cite{frohling2025personas,feng2025crowdllm,labarbera2025impersonating}, raising questions about how persona-based evaluation should be interpreted. A separate line of research models personas through psychological traits or demographic attributes~\cite{jiang2023personality,personalllm2024,sorokovikova2024}. In relevance assessment specifically, personality-conditioned prompts have been reported to yield modest alignment improvements~\cite{chen2026personalityjudge}, though demographic and persona prompting more broadly often produce inconsistent effects and limited controllability~\cite{zheng2024persona,culpepper2025demographic,sclar2023prompt,zhan2024lexical}. In contrast to this body of work, which primarily models personas through demographic or psychological traits, we adopt \emph{task-oriented} assessor perspectives and analyze how these perspectives propagate to downstream IR evaluation stability and retrieval-system sensitivity.

\section{Methodology and Experimental Setting}
This section describes our methodology and experimental setup. Our goal is to use persona-conditioned judging as a diagnostic probe for evaluating how changes in assessor perspective affect relevance labels and downstream IR evaluation outcomes.\footnote{Code is publicly available at \url{https://osf.io/5sf2z/overview?view_only=e57aef041bb142d59722c44ef4f5682b}}
\subsection{Datasets and Models}
We conduct experiments on two benchmark datasets commonly used in recent IR evaluation with LLM judges: TREC Deep Learning 2020 (DL20)~\cite{CraswellMMYC20} and the TREC Retrieval-Augmented Generation track 2024 (RAG24)~\cite{trec2024rag}. DL20 contains 54 predominantly short, factual, and definitional queries, whereas RAG24 contains 86 longer, more open-ended queries that often involve explanatory, contextual, or subjective information needs. Both datasets employ graded relevance labels, and we include all available submitted system runs for each dataset in our system-level evaluation. We evaluate six LLM backbones covering a range of model families and scales, including GPT-4o and GPT-4o-mini, LLaMA-3.1-70B and LLaMA-3.1-8B, and Qwen-2.5-72B and Qwen-2.5-7B. We use instruction-tuned conversational variants of the open-weight models~\cite{ouyang2022training}, which are well suited to our in-context prompting approach for eliciting distinct assessor perspectives during relevance judgment. All judgments are generated with temperature set to 0 to reduce sampling variability across repeated assessor conditions.
\subsection{Assessor Personas}
In this work, we model LLM judges using task-oriented assessor personas. We use the term persona to denote an explicit evaluation role that reflects a particular relevance assessment perspective, rather than psychological personality traits or demographic attributes. This distinction is central to our study, which focuses on how relevance is judged, rather than who the judge is. 
\subsubsection{Assessor Roles}
We define five assessor roles that emphasize different relevance perspectives: (i) query-aligned interpretation, (ii) domain expertise, (iii) contrastive or orthogonal judgment, (iv) evidence verification, and (v) a global assessor perspective.
\paragraph{Query-Aligned Assessor.}
The Query-Aligned assessor focuses on the user's information need, favoring documents that semantically align with the inferred query intent. This role is instantiated separately for each query to reflect query-specific interpretation.
\paragraph{Domain-Expert Assessor.}
The Domain-Expert assessor, also referred to as Domain, emphasizes topical expertise and domain-specific relevance criteria. To instantiate this role consistently across datasets, we construct a shared nine-domain taxonomy and assign each query to exactly one domain (Table~\ref{tab:domains}). The mapping was produced by two independent annotators, with disagreements resolved through discussion after measuring inter-annotator agreement using Cohen's $\kappa$ ($0.86$). This mapping is used for Domain persona retrieval and domain-level aggregation.
\begin{table}[t]
\mycaption{Domain taxonomy used for Domain-Expert assessor personas. Each query is assigned to exactly one domain.}
\label{tab:domains}
\centering
\footnotesize
\setlength{\tabcolsep}{2pt} 
\begin{tabular}{@{}lr lr@{}} 
\toprule
\textbf{Domain} & \textbf{Abbr.} & \textbf{Domain} & \textbf{Abbr.} \\
\midrule
Health and Medicine             & H\&M & Environment and Earth Sciences & ENV \\
Law, Policy, and Government     & LPG  & Current Affairs and History    & CAH \\
Science and Technology          & S\&T & Education and Humanities       & E\&H \\
Business and Economics          & B\&E & Arts, Media, and Culture       & AMC \\
General Knowledge and Reasoning & GKR  &                                &     \\
\bottomrule
\end{tabular}
\end{table}
\begin{table*} [htbp]
\mycaption{Illustrative persona examples for the RAG24 query \textit{``should teachers notify parents about state testing?''} across both {\tt{PersonaHub}} and {\tt{USPersona}} data. Personas taken from {\tt{USPersona}} tend to be longer and more skill-oriented.}
\label{tab:persona_examples}
\centering
\footnotesize
\setlength{\tabcolsep}{3pt}
\renewcommand{\arraystretch}{0.96}
\begin{tabularx}{\textwidth}{llX}
\toprule
\textbf{Source} & \textbf{Role} & \textbf{Example Persona} \\
\midrule

\multirow{4}{*}{\tt PersonaHub}
& Query
& A single parent advocating for education reform based on their child's stress from excessive testing. \\[1.2ex]

& Domain
& A sociology professor researching the impact of incorporating humanities in STEM education on student success and creativity. \\[1.2ex]

& Orthogonal
& A highly sought-after designer known for creating chic and trendy nightclub interiors. \\ \addlinespace[1.2ex]

\midrule

\multirow{10}{*}{\tt USPersona}
& Query
& A persona with the following skills: Early childhood education, STEM curriculum integration for preschool, Structured daily routine design, Low-stimulus classroom environment creation, Individualized learning plan development, Observational child assessment, Educational technology utilization, Hands-on science experiment facilitation, Literacy foundation teaching, Emotional regulation techniques for children, Parent communication and consultation, Curriculum adaptation for anxiety-prone learners. \\ \addlinespace[1.2ex]

& Domain
& A persona with the following skills: Interdisciplinary curriculum development, Instructional technology integration, Public speaking and lecturing, Classroom management, Conflict resolution, Community outreach and partnership building, Research and academic writing, Flexible lesson planning, Mentorship and coaching, Cultural competency in humanities education. \\ \addlinespace[1.2ex]

& Orthogonal
& A persona with the following skills: Vendor negotiation, Strategic sourcing, Cost analysis and budgeting, Supply chain risk assessment, Market research, Creative procurement problem solving, Sustainability sourcing, Contract drafting, Data analytics with Excel and Power BI, Use of e-procurement platforms (SAP Ariba, Coupa), Improvisational logistics planning, Relationship management with suppliers, Inventory forecasting, Process optimization, Cross-functional collaboration. \\ \addlinespace[1.2ex]

\bottomrule
\end{tabularx}
\vspace{-2mm}
\end{table*}
\paragraph{Orthogonal Assessor.}
The Orthogonal assessor introduces a contrastive perspective by intentionally deviating from dominant interpretations of query intent. This role is designed to surface assessor sensitivity under alternative assessor framing, rather than to simulate an incorrect or random judge. Orthogonal assessors are instantiated on a per-query basis using semantically dissimilar persona retrieval to encourage contrastive assessor framing.
\paragraph{Evidence-Verification Assessor.}
The Evidence-Verification assessor emphasizes factual correctness, evidential support, and source credibility in relevance judgments. This role targets RAG-style evaluation scenarios, where documents may be topically related yet poorly supported or unreliable, and therefore penalizes speculative or unsubstantiated claims even when topical alignment is strong. Unlike query-specific personas, this assessor is defined through a fixed prompt to explicitly isolate a trust-oriented relevance dimension. We define this persona as: ``An assessor who prioritizes factual correctness, evidential support, and source credibility when judging relevance. Favor well-supported and verifiable content, and penalize speculative or unsubstantiated claims, even when topically related.''
 \paragraph{Global Assessor Persona (GAP)}
Following prior work that uses global assessor impersonation~\cite{thomas2024large}, we define GAP as a professional search-quality rater applying standard web-search relevance guidelines. GAP is expressed through a fixed system prompt and is applied uniformly across all queries and datasets. Compared to the UMBRELA default, GAP makes the assessor perspective explicit, providing a stable query-independent reference for comparison with query-specific persona-conditioned judgments. We define this persona as: ``A professional search quality rater evaluating relevance according to standard web search guidelines.''
\subsubsection{Persona Sources}
We instantiate the Query-Aligned, Domain-Expert, and Orthogonal roles from two complementary persona sources in order to compare abstract task-oriented personas with skill-grounded professional personas.

The first source is PersonaHub~\cite{ge2024scaling}, which provides synthetic persona descriptions designed to support diversity in annotation and evaluation tasks~\cite{frohling2025personas,bernardelle2025mapping,civelli2025ideology,bernardelle2025political,civelli2025impact,bernardelle-etal-2026-subdata}. PersonaHub personas are abstract and broad in scope, often encoding viewpoints, occupations, or experiential backgrounds intended to diversify interpretation.

The second source, which we refer to as USPersona, is derived from the NVIDIA Nemotron-Personas-USA collection~\cite{nvidia/Nemotron-Personas-USA}. Unlike PersonaHub, USPersona provides occupationally grounded profiles associated with explicit skills and expertise. To preserve task-oriented assessor framing rather than demographic or psychological conditioning, we restrict retrieval to professional-description and skill-related fields while excluding demographic, cultural, and lifestyle attributes. Retrieval is therefore performed over skill-based descriptions of the form \emph{``A person with the following skills: [skills list]''}, enabling construction of assessor roles grounded in concrete expertise rather than identity characteristics.

For both PersonaHub and USPersona, persona selection follows the same retrieval procedure. Persona descriptions are encoded using \texttt{all-MiniLM-L6-v2}, and cosine similarity is computed between persona embeddings and either query or domain representations. At each selection step, we retrieve the top three candidate personas for inspection and retain the highest-ranked one. Query-Aligned and Domain-Expert personas are selected based on high similarity to the query or domain context, while Orthogonal personas are selected to be intentionally dissimilar. This process allows different queries to activate different assessor profiles, reflecting structured variation in interpretation rather than fixed assessor identity. 
\subsection{UMBRELA Baseline}
The UMBRELA assessor corresponds to the default UMBRELA relevance-judging configuration~\cite{upadhyay2024umbrelasystem} and serves as the primary baseline. We use UMBRELA as a fixed judging reference. This allows us to measure how explicit assessor-role conditioning changes relevance judgments and downstream system rankings relative to a standard LLM judging setup.

\subsection{Summary-Based Judging}
\label{sec:summary_judging}
Across both datasets, each query is evaluated under nine judging conditions: PersonaHub Query, PersonaHub Domain, PersonaHub Orthogonal, USPersona Query, USPersona Domain, USPersona Orthogonal, Evidence-Verification, GAP, and UMBRELA. Since each query--document pair is judged repeatedly across assessor roles, persona sources, and model backbones, the total number of LLM evaluations grows rapidly. We therefore adopt summary-based judging, following prior work showing that concise document summaries can preserve agreement and system-level ranking stability while substantially reducing evaluation cost~\cite{mohtadi2026summarization}.

For all experiments, each document is summarized once into an approximately 80-token summary, and the resulting summary is reused across all assessor personas, persona sources, model backbones, and evaluation settings. Document summaries are generated using the prompt template and generation settings released with the prior summary-based judging study~\cite{mohtadi2026summarization}; following that setup, we use GPT-4o as the summarization model.
This design enables controlled experimentation by fixing the document representation and varying only the assessor perspective. As a result, observed differences in relevance judgments are less likely to arise from document length, context variability, or repeated summarization. Summary reuse is particularly important in our setting because each query--document pair is evaluated multiple times under different assessor roles and LLM configurations.
\subsection{Evaluation Framework}
We adopt the UMBRELA judging prompt as the underlying relevance-judging framework and vary only the assessor-role instruction. For persona-conditioned runs, the UMBRELA prompt~\cite{upadhyay2024umbrelasystem} is prefixed with the assessor instruction ``You are acting as \{persona\}.'' This design keeps the judging template fixed, allowing observed differences to be attributed to persona conditioning rather than prompt-template changes. Our evaluation considers three complementary levels.
\paragraph{Judgment-level agreement.}
We measure how persona-conditioned judgments differ from UMBRELA and human judgments. Because relevance labels are ordinal, we report quadratic-weighted Cohen's $\kappa$~\cite{cohen1968weighted}.
Quadratic weighting accounts for the ordinal structure of relevance labels by penalizing distant disagreements more strongly than adjacent-grade shifts. This allows us to distinguish severe relevance inversions from local grade shifts induced by persona conditioning.
\paragraph{System ranking stability.}
At the system level, we first compute retrieval effectiveness using NDCG@10~\cite{jarvelin2002cumulated}, following standard TREC Deep Learning practice~\cite{CraswellMMYC20}. For each assessor configuration, including UMBRELA, we derive system-level effectiveness scores from LLM-derived relevance labels and rank submitted runs accordingly. We then compare each LLM-induced system ranking with the human-derived ranking using Kendall's $\tau$, a standard measure of pairwise ranking consistency widely used to study sensitivity to assessor variation~\cite{kendall1938new,gera2024justrank}, and Rank-Biased Overlap (RBO) with persistence parameter $\phi=0.9$, which emphasizes agreement among top-ranked systems~\cite{webber2010similarity}.
\paragraph{System-level sensitivity.}\label{para:system-level-sensitivity}
In addition to global ranking stability, we measure localized system impact through rank-displacement analysis. For each retrieval system $s$ and persona condition $p$, we compute the rank shift relative to UMBRELA:
\[
\Delta r(s,p) = r_p(s) - r_{\mathrm{UMB}}(s),
\]
where negative values indicate that a system moves up under persona-conditioned judging and positive values indicate that it moves down. We summarize system sensitivity as:
\[
\mathrm{Sensitivity}(s) = \frac{1}{|P|}\sum_{p \in P} |\Delta r(s,p)|,
\]
where $P$ denotes the set of non-UMBRELA persona conditions. Mean absolute rank shift is first computed across retrieval systems for each persona condition and model, and then averaged across models for dataset-level reporting. To assess the robustness of model sensitivity estimates, we additionally compute 95\% bootstrap confidence intervals by resampling retrieval systems with replacement 1,000 times within each dataset--model pair (random seed: 42).

We additionally report maximum rank displacement, affected-system counts, signed NDCG@10 shifts, and direction consistency across models. Maximum rank displacement captures the largest observed rank movement under persona conditioning. Affected-system counts measure how many systems change rank relative to UMBRELA, while signed NDCG@10 shifts indicate whether systems improve or degrade under a persona condition. Direction consistency measures whether systems move consistently upward or downward across models. Together, these analyses identify retrieval systems whose evaluation outcomes are most sensitive to changes in assessor perspective and distinguish global ranking stability from localized system movement.
\section{Results and Analysis}
The results show how persona conditioning affects LLM judgments at three levels: relevance-label agreement, downstream system-ranking stability, and localized system sensitivity.
\begin{figure*}[!t]
    \centering
    \includegraphics[width=0.85\textwidth]{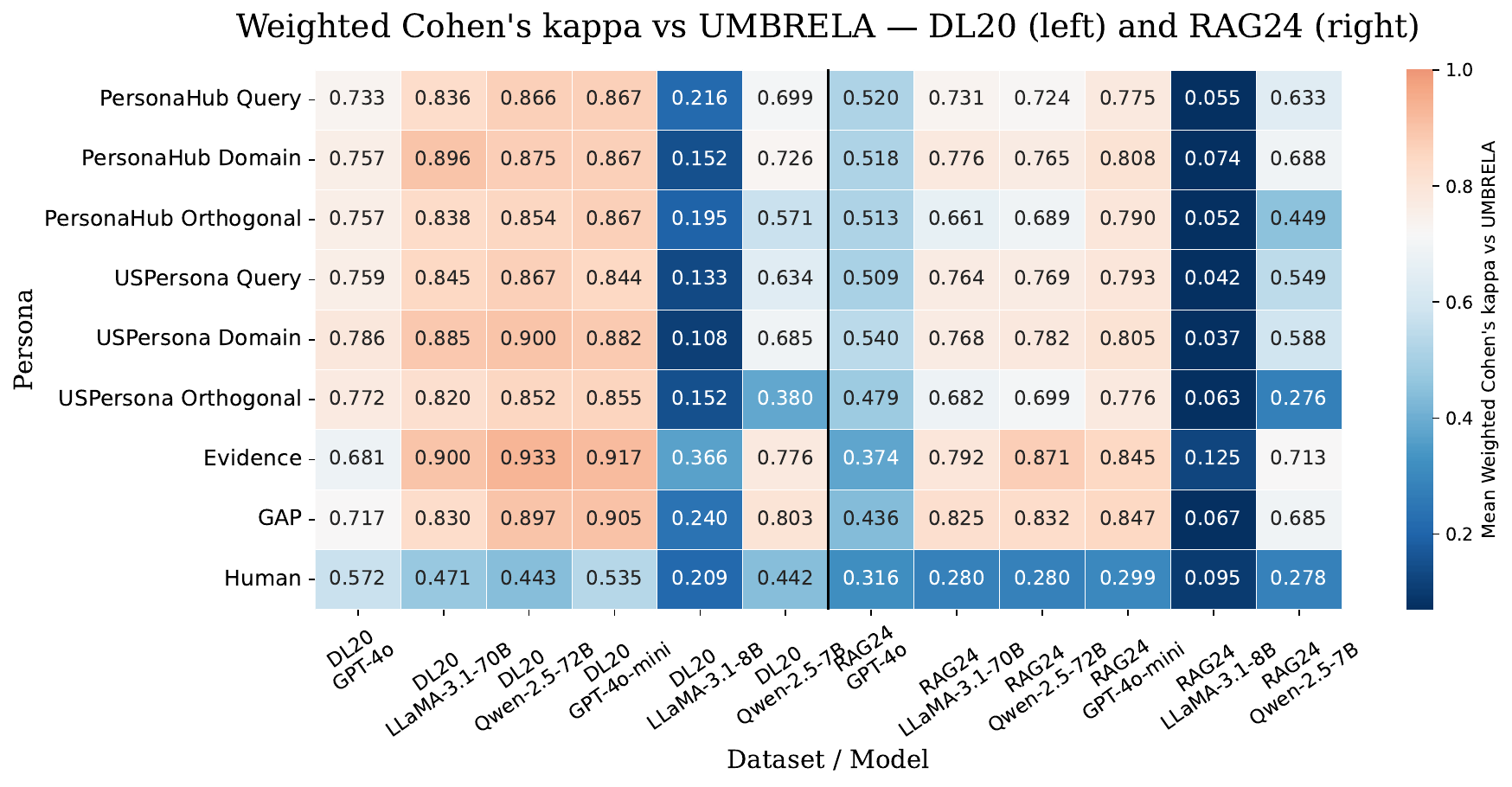}
    \vspace{-2mm}
    \mycaption{Persona-level agreement with UMBRELA judging. Heatmap values show the mean quadratic-weighted Cohen's $\kappa$ agreement between persona-conditioned and UMBRELA judgments, averaged over all queries.}
    \label{fig:kappa-vs-vanilla}
    \vspace{-2mm}
\end{figure*}
\subsection{RQ1: Effects of Persona Conditioning on Relevance Judgments}
We first examine how assessor perspectives influence relevance judgments relative to the default UMBRELA judging configuration, considering agreement with UMBRELA and agreement with human judgments across datasets and model families.
\subsubsection{Agreement with UMBRELA Judging} \label{sec:rq1_umbrela}
Figure~\ref{fig:kappa-vs-vanilla} reports the mean quadratic-weighted Cohen's $\kappa$ between persona-conditioned and UMBRELA judgments, averaged across queries for each dataset and model. Higher values indicate stronger agreement with the baseline UMBRELA configuration, whereas lower values indicate larger persona-induced deviations.

Overall, persona-conditioned judgments remain close to UMBRELA for most models. Agreement is highest for the high-capacity LLaMA-3.1-70B and Qwen-2.5-72B, with mid-capacity GPT-4o-mini exhibiting similarly stable behavior. GPT-4o is a notable exception: despite being a high-capacity model, it maintains only moderate-to-high agreement on DL20 and drops further on RAG24, especially under Evidence and GAP personas, indicating greater sensitivity to assessor perspective on the more interpretive dataset. In contrast, smaller models, particularly LLaMA-3.1-8B and Qwen-2.5-7B, show substantially lower agreement and greater variability across assessor roles.

Although role-level patterns vary across models and datasets, a consistent diagnostic trend emerges in the heatmap: global assessor perspectives such as Evidence and GAP often remain close to UMBRELA, whereas Orthogonal personas are more likely to induce larger deviations. Query and Domain personas generally fall between these two extremes.
\subsubsection{Agreement with Human Judgments}
To assess how persona conditioning affects agreement with human judgments, we perform a per-query win-rate analysis relative to UMBRELA. For each persona source, assessor role, and model, we compare persona-conditioned and UMBRELA judgments on the same query. A persona is counted as a win for a query if its quadratic-weighted Cohen's $\kappa$ agreement with human judgments is higher than that of the corresponding UMBRELA judge. The win rate (W\%) is then computed as the proportion of queries for which the persona achieves higher agreement with human judgments than UMBRELA. We also report the mean agreement difference ($\bar{\Delta}$), computed as the average per-query difference in weighted $\kappa$ between the persona-conditioned assessor and UMBRELA. Positive values of $\bar{\Delta}$ indicate increased alignment with human judgments relative to UMBRELA, while negative values indicate decreased alignment.
\input{win_rate_table}
Table~\ref{tab:winrate_persona} reports W\% and $\bar{\Delta}$ for each persona source, dataset, model, and assessor role. Because each comparison is made against the same model under UMBRELA judging, the analysis should be interpreted as measuring persona-induced movement toward or away from human judgments rather than absolute judgment quality.

Persona conditioning does not uniformly improve agreement with human judgments. On DL20, several mid- and high-capacity models obtain win rates above 50\% for selected assessor roles, often with small positive $\bar{\Delta}$ values. This pattern is most visible for GPT-4o-mini and Qwen-2.5-72B, while LLaMA-3.1-70B shows mixed but occasionally positive shifts. In contrast, LLaMA-3.1-8B exhibits consistently low win rates and strongly negative agreement shifts across nearly all roles and datasets, indicating high evaluator instability under persona conditioning.

Dataset-level differences are also apparent. Persona-conditioned judging produces more positive or near-neutral shifts on DL20, whereas RAG24 shows weaker and less consistent gains, with $\bar{\Delta}$ values often close to zero or negative. GPT-4o shows a different pattern among high-capacity models: although it maintains reasonable agreement with UMBRELA, its win rates against human judgments are lower than most other mid- and high-capacity models, and all $\bar{\Delta}$ values on RAG24 are negative. This suggests that persona conditioning moves GPT-4o away from human judgments on the more interpretive RAG24 queries rather than increasing agreement.

The win-rate results therefore show that persona-induced changes do not translate into uniform gains in human agreement. Consistent with the UMBRELA-agreement heatmap, the direction and magnitude of these changes depend on the interaction between model capacity, dataset characteristics, and assessor role. Thus, persona conditioning acts as a controlled source of evaluator variation, revealing where judgments are sensitive to assessor perspective.
\subsection{RQ2: Abstract vs Skill-Grounded Personas}
\label{sec:rq2}
We compare abstract PersonaHub personas with skill-grounded USPersona profiles under matched Query, Domain, and Orthogonal assessor roles. As shown in Figure~\ref{fig:kappa-vs-vanilla} and Table~\ref{tab:winrate_persona}, the two persona sources produce broadly similar agreement and win-rate patterns, with no uniformly stronger source across datasets or models. Differences are concentrated in particular role--model combinations: USPersona sometimes amplifies positive shifts, such as Domain with LLaMA-3.1-70B on DL20, but can also amplify negative shifts, such as Orthogonal with Qwen-2.5-7B on RAG24. 

Overall, the source of the persona description affects the magnitude of assessor-perspective effects, but does not fundamentally change the stability pattern observed across roles, datasets, and model families. One possible explanation is that both PersonaHub and USPersona operationalize task-oriented assessor framing through the same retrieval and prompting pipeline. Although the two sources differ in abstraction level and specificity, both guide the model toward similar relevance-assessment perspectives. This suggests that persona source mainly modulates the magnitude of assessor-perspective effects, whereas the role instruction and model determine the dominant stability pattern.
\begin{figure*}[ht]
  \centering
  \includegraphics[width=0.95\textwidth]{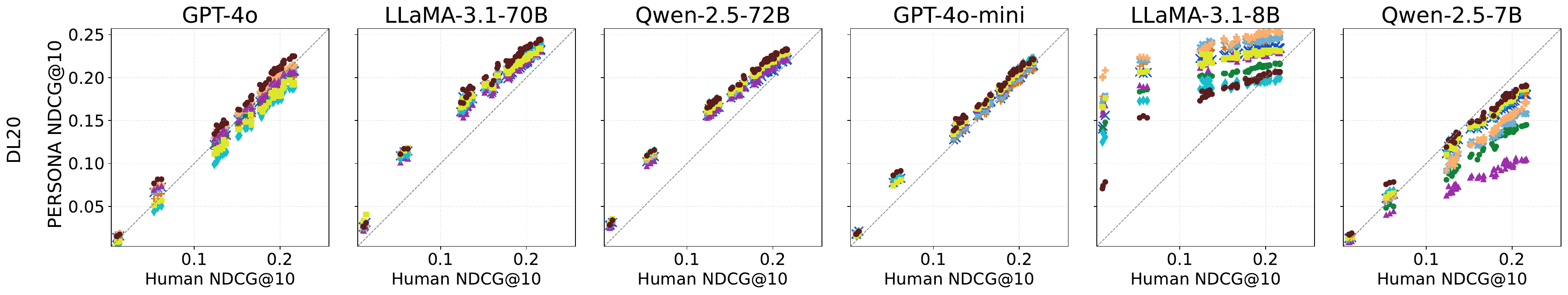}\par\vspace{1.5mm}
 \includegraphics[width=0.95\textwidth]{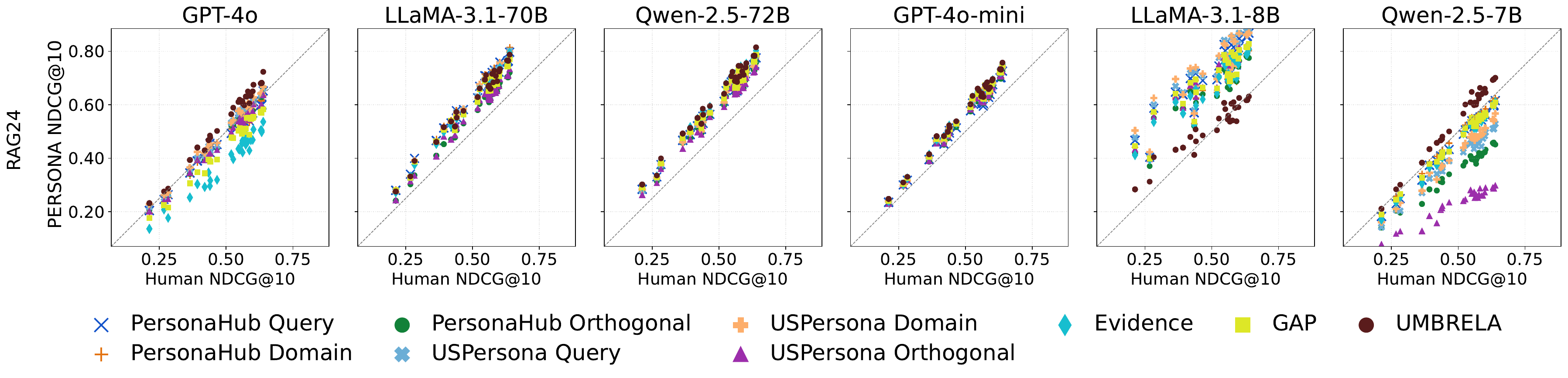}\par
  \mycaption{Human- vs.\ LLM-derived retrieval effectiveness under persona-conditioned judging for DL20 (top) and RAG24 (bottom). Each point represents a retrieval system scored using mean NDCG@10 across queries.}
  \label{fig:ndcg-scatter}
\end{figure*}
\subsection{RQ3: Stability of System Rankings}
\input{system_stability_tau_rbo}
We next evaluate whether persona-conditioned judgments preserve downstream IR system rankings derived from human relevance judgments across datasets, assessor perspectives, and model families.
\subsubsection{System Effectiveness under Persona-Conditioned Judging}
Figure~\ref{fig:ndcg-scatter} reports system-level NDCG@10 scores computed using persona-conditioned relevance judgments on DL20 (top row) and RAG24 (bottom row), plotted against human-derived effectiveness. Each point corresponds to a submitted retrieval system under a particular assessor perspective. The diagonal line ($y=x$) indicates perfect agreement with human-derived effectiveness, and deviations from the diagonal reflect persona-induced changes in the evaluation.

Across most models and datasets, the scatter plots exhibit a strong monotonic relationship with the human baseline, indicating that systems ranked highly under human judgments generally remain highly ranked under persona-conditioned judging. For high-capacity models, persona-conditioned effectiveness estimates tend to form approximately parallel bands around the diagonal rather than large-scale crossing patterns. The main visible exception is GPT-4o on RAG24, where several assessor perspectives pull effectiveness estimates below the diagonal, consistent with the lower UMBRELA agreement observed in Section~\ref{sec:rq1_umbrela}. This suggests that assessor perspectives primarily introduce systematic shifts in scoring behavior, such as stricter or more lenient effectiveness estimation, while largely preserving relative system ordering.

Across assessor roles, Query and Domain perspectives often overlap closely, whereas Orthogonal perspectives tend to produce larger shifts in effectiveness estimates. Evidence and GAP usually remain closer to the human baseline for stronger models, although this behavior varies across datasets and model families. The persona-source comparison shows the same pattern observed at the judgment level (Section~\ref{sec:rq2}): broadly similar behavior, with differences concentrated in role--model combinations rather than as a consistent source-level effect.

Overall, the scatter plots indicate that persona conditioning generally preserves the global structure of system effectiveness while introducing assessor-dependent shifts in evaluation scale. Large-scale rank inversions remain uncommon for stronger models, whereas lower-capacity models exhibit greater assessor sensitivity and more pronounced persona-dependent variation.
\subsubsection{Ranking Consistency Across Models}
To assess ranking stability under persona-conditioned judging, we compute system rankings from LLM-derived NDCG@10 scores and compare them against rankings derived from human relevance judgments. We report Kendall’s $\tau$ and Rank-Biased Overlap (RBO, $\phi=0.9$), where Kendall’s $\tau$ measures global rank-order consistency and RBO emphasizes agreement among top-ranked systems. Table~\ref{tab:system_stability_tau_rbo} summarizes ranking agreement across datasets, assessor perspectives, and models.

Overall, high-capacity models generally preserve strong ranking agreement with human-derived system rankings across most assessor perspectives. LLaMA-3.1-70B, Qwen-2.5-72B, GPT-4o-mini, and GPT-4o usually achieve high Kendall’s $\tau$, indicating that persona conditioning rarely disrupts global system ordering for stronger models. In contrast, lower-capacity models, particularly LLaMA-3.1-8B, exhibit reduced ranking stability and substantially lower RBO values under several assessor perspectives, indicating greater instability among top-ranked systems.

Across datasets, ranking stability does not follow a uniform DL20--RAG24 pattern. Kendall's $\tau$ remains high for stronger models on both datasets, but RBO varies more sharply across assessor perspectives and models. This suggests that persona conditioning often preserves global system ordering while still affecting the relative ordering of top-ranked systems. The effect is most visible for smaller models, where some assessor perspectives produce substantial top-rank volatility despite moderate global rank agreement.

Persona effects are role-dependent. Domain, Orthogonal, and Evidence perspectives sometimes achieve ranking agreement comparable to or higher than UMBRELA, particularly for high-capacity models, but this pattern is not consistent across datasets, models, or stability metrics. No single assessor perspective is uniformly most stable. Instead, ranking stability depends on the interaction between model capacity, dataset, metric, and assessor perspective. The persona-source comparison shows the same pattern as at the judgment level (Section~\ref{sec:rq2}): persona source has a secondary effect relative to model capacity, dataset, and assessor role.

\subsection{RQ4: System-Level Sensitivity}
Although global ranking agreement remains relatively stable, 
individual retrieval systems may still move substantially under persona-conditioned judging. We therefore analyze rank displacement relative to UMBRELA to identify which assessor perspectives and retrieval systems are most evaluator-sensitive.

\subsubsection{Rank-Displacement Sensitivity Analysis}
To quantify localized movement, we analyze system-level rank displacement relative to UMBRELA using the sensitivity measures defined in Section~\ref{para:system-level-sensitivity}. Table~\ref{tab:persona_sensitivity} summarizes rank displacement by assessor perspective. USPersona Orthogonal yields the highest mean absolute rank displacement on both DL20 (2.31) and RAG24 (2.66), indicating that contrastive skill-grounded personas act as strong evaluator-sensitivity probes. PersonaHub Orthogonal also produces elevated displacement on RAG24 (2.22) but remains moderate on DL20 (1.67), so the Orthogonal role is not uniformly the most disruptive perspective. Domain perspectives consistently induce among the smaller shifts on both datasets (USPersona Domain: 1.46 on DL20, 1.55 on RAG24), while GAP behaves inconsistently---lowest on RAG24 (1.43) but third-highest on DL20 (2.11). RAG24 shows larger score perturbations overall, especially in mean absolute $\Delta$NDCG@10, peaking at 0.052 for USPersona Orthogonal compared to 0.013 on DL20. This suggests greater assessor sensitivity in broader and more interpretive retrieval settings. Despite these perturbations, the average score changes remain bounded, indicating localized movement rather than large-scale effectiveness collapse.
\begin{table}[!h]
\centering
\mycaption{System-level sensitivity relative to UMBRELA. Mean and max $|\Delta r|$ report average and largest rank displacement; mean $|\Delta\mathrm{NDCG}@10|$ reports average score perturbation. Bold indicates the largest mean $|\Delta r|$ per dataset.}
\label{tab:persona_sensitivity}
\small
\setlength{\tabcolsep}{4pt}
\renewcommand{\arraystretch}{1.08}
\begin{tabular*}{\columnwidth}{@{\extracolsep{\fill}}llccc@{}}
\toprule
\textbf{Dataset} & \textbf{Persona} &
\textbf{Mean} &
\textbf{Max} &
\textbf{Mean} \\
&
&
\textbf{$|\Delta r|$} &
\textbf{$|\Delta r|$} &
\textbf{$|\Delta\mathrm{NDCG}@10|$} \\
\midrule
DL20
& PersonaHub Query      & 2.05 & 17 & 0.003 \\
& USPersona Query       & 1.99 & 20 & 0.003 \\
& PersonaHub Domain     & 1.75 & 17 & 0.001 \\
& USPersona Domain      & 1.46 & 11 & 0.002 \\
& PersonaHub Orthogonal & 1.67 & 13 & 0.010 \\
& \textbf{USPersona Orthogonal} & \textbf{2.31} & 27 & \textbf{0.013} \\
& Evidence              & 2.21 & \textbf{32} & 0.007 \\
& GAP                   & 2.11 & 20 & 0.002 \\
\midrule
RAG24
& PersonaHub Query      & 1.83 & 17 & 0.004 \\
& USPersona Query       & 2.29 & 17 & 0.003 \\
& PersonaHub Domain     & 1.96 & 17 & 0.000 \\
& USPersona Domain      & 1.55 & 16 & 0.005 \\
& PersonaHub Orthogonal & 2.22 & \textbf{20} & 0.035 \\
& \textbf{USPersona Orthogonal} & \textbf{2.66} & \textbf{20} & \textbf{0.052} \\
& Evidence              & 1.97 & 17 & 0.016 \\
& GAP                   & 1.43 & 18 & 0.004 \\
\bottomrule
\end{tabular*}
\vspace{-2mm}
\end{table}
To examine the role of model capacity, we additionally aggregate system sensitivity by model. Table~\ref{tab:judge_model_sensitivity} reports mean system-rank displacement with 95\% bootstrap confidence intervals, computed by resampling retrieval systems with replacement 1,000 times within each dataset--model pair.

Model capacity strongly modulates sensitivity to persona conditioning, although the pattern is not strictly monotonic across datasets. LLaMA-3.1-8B produces the largest average system-rank displacement on both datasets, with bootstrap intervals that remain clearly higher than those of the higher-capacity judge models. Qwen-2.5-7B also exhibits elevated instability, while Qwen-2.5-72B and GPT-4o-mini remain comparatively stable. LLaMA-3.1-70B is highly stable on DL20 but more sensitive on RAG24, indicating that dataset characteristics also shape persona-induced rank displacement.

\begin{table}[t]
\centering
\mycaption{Model sensitivity to persona-conditioned evaluation. Mean $|\Delta r|$ reports average absolute system-rank displacement with 95\% bootstrap confidence intervals from resampling retrieval systems. Max $|\Delta r|$ reports the largest observed movement for any retrieval system.}
\label{tab:judge_model_sensitivity}
\small
\setlength{\tabcolsep}{4pt}
\renewcommand{\arraystretch}{1.08}
\begin{tabular*}{\columnwidth}{@{\extracolsep{\fill}}llcc@{}}
\toprule
\textbf{Dataset} & \textbf{Model} &
\textbf{Mean $|\Delta r|$ [95\% CI]} &
\textbf{Max $|\Delta r|$} \\
\midrule
DL20  & LLaMA-3.1-8B  & \textbf{5.19 [4.60, 5.79]} & \textbf{32} \\
      & Qwen-2.5-7B   & 2.15 [1.79, 2.54] & 15 \\
      & GPT-4o        & 1.56 [1.21, 1.96] & 12 \\
      & GPT-4o-mini   & 1.10 [0.87, 1.37] & 8 \\
      & Qwen-2.5-72B  & 0.86 [0.67, 1.07] & 8 \\
      & LLaMA-3.1-70B & 0.80 [0.64, 0.98] & 5 \\
\midrule
RAG24 & LLaMA-3.1-8B  & \textbf{4.17 [3.48, 4.88]} & \textbf{20} \\
      & LLaMA-3.1-70B & 2.21 [1.59, 2.91] & 12 \\
      & Qwen-2.5-7B   & 2.01 [1.61, 2.53] & 19 \\
      & GPT-4o        & 1.80 [1.46, 2.20] & 17 \\
      & GPT-4o-mini   & 0.99 [0.69, 1.33] & 10 \\
      & Qwen-2.5-72B  & 0.75 [0.52, 1.01] & 10 \\
\bottomrule
\end{tabular*}
\vspace{-2mm}
\end{table}

These results show that persona effects are concentrated rather than uniformly distributed across rankings. Persona conditioning rarely disrupts global rankings for high-capacity models, but it exposes localized evaluator-sensitive systems and amplifies instability in lower-capacity evaluators.
\begin{table*}[h]
\centering
\mycaption{Representative most evaluator-sensitive retrieval systems under persona-conditioned judging. The reported sensitivity ranges are computed over the listed systems and measured as mean absolute rank displacement relative to UMBRELA across persona settings.}
\label{tab:sensitive_systems}
\small
\setlength{\tabcolsep}{5pt}
\renewcommand{\arraystretch}{1.1}
\begin{tabularx}{\textwidth}{llccX}
\toprule
\textbf{Dataset} &
\textbf{Representative Systems} &
\textbf{Mean $|\Delta r|$} &
\textbf{Max $|\Delta r|$} &
\textbf{Interpreted System Type} \\
\midrule
DL20
& \texttt{bigIR-DCT-T5-F}, \texttt{pash\_r1}, \texttt{fr\_pass\_roberta}
& 3.69--4.19
& 18--27
& Transformer-based neural ranking/reranking systems \\
\midrule
RAG24
& \texttt{iiia\_standard\_*}, \texttt{ielab-*}
& 4.31--8.75
& 19--20
& Retrieval-augmented and generation-oriented pipelines \\
\bottomrule
\end{tabularx}
\vspace{-2mm}
\end{table*}
\subsubsection{Sensitivity Across System Types}
Next, we examine whether assessor sensitivity is concentrated in particular types of retrieval systems. Following prior TREC Deep Learning analyses, which distinguish reranking and neural language model runs~\cite{CraswellMMYC20}, and TREC RAG work that frames systems around retrieval-augmented generation pipelines~\cite{pradeep2025ragnarok}, we identify the most affected runs using broad system-type descriptions rather than official track labels. These descriptions are derived from official TREC run descriptions and associated system papers, and are used only to interpret sensitivity patterns rather than as additional experimental variables.

Table~\ref{tab:sensitive_systems} reports representative systems with the highest evaluator sensitivity in each dataset. The reported Mean $|\Delta r|$ and Max $|\Delta r|$ ranges are computed over the listed systems, where $\Delta r$ denotes rank displacement relative to UMBRELA across persona settings. On DL20, several of the most evaluator-sensitive runs are transformer-based neural ranking and reranking systems, including \texttt{bigIR-DCT-T5-F}, \texttt{fr\_pass\_roberta}, and \texttt{pash\_r1}. These systems exhibit larger rank displacement than typical runs under persona-conditioned judging.

On RAG24, the most evaluator-sensitive runs we identify are retrieval-augmented and generation-oriented pipelines, including \texttt{iiia\_standard\_*} and \texttt{ielab-*}. These runs exhibit larger average displacement and comparable or higher maximum rank shifts than the most sensitive DL20 runs, suggesting that RAG-style settings may amplify assessor sensitivity. This pattern is also consistent with the earlier model analysis: the largest perturbations are typically associated with smaller models such as LLaMA-3.1-8B and Qwen-2.5-7B, whereas higher-capacity models produce more bounded movement.

We further examined whether persona-induced rank movement is directionally consistent across models. Most systems exhibit mixed or unchanged movement directions, particularly on RAG24, where no system moves consistently up or down across all models. On DL20, only 7 out of 472 system--persona pairs show consistent directional behavior. For example, \texttt{fr\_pass\_roberta} moves upward under both PersonaHub Orthogonal and USPersona Orthogonal perspectives across all six models, with mean rank shifts of $-4.17$ and $-4.83$, respectively. This suggests that system sensitivity emerges from the interaction between assessor perspective and model capacity rather than being a fixed property of the retrieval system alone.

Overall, these results suggest that persona conditioning acts as a diagnostic stress test that exposes evaluator-sensitive retrieval architectures rather than uniformly disrupting system rankings.



%
\section{Conclusions and Implications}
This work examined how assessor perspectives influence LLM-based relevance judgments and downstream IR evaluation. Across agreement analysis, ranking stability, and system-sensitivity analysis, a consistent set of patterns emerges.

First, persona conditioning systematically alters relevance judgments, but the magnitude and structure of these effects depend strongly on model capacity and dataset characteristics. High-capacity models generally maintain strong agreement with both UMBRELA and human judgments, while lower-capacity models exhibit substantially greater instability under changes in assessor perspective. This indicates that assessor conditioning is meaningful only when the underlying model can reliably support the imposed evaluator constraints; otherwise, persona conditioning primarily amplifies judgment variability.

Second, persona effects are localized rather than globally disruptive. Although persona-conditioned judgments can induce noticeable shifts in effectiveness estimates and individual system ranks, overall ranking agreement with human evaluation remains relatively stable for stronger models. The scatter plots and rank-correlation analyses show that persona conditioning usually preserves the global structure of system rankings while producing bounded movement among individual systems, particularly near the top of the ranking. This suggests that assessor perspectives mainly change scoring behavior and local rank positions, rather than producing widespread rank inversions.

Third, assessor sensitivity is not uniformly distributed across retrieval systems. Persona-induced perturbations concentrate on particular systems and system types: among the most evaluator-sensitive runs are transformer-based neural ranking/reranking systems on DL20 and retrieval-augmented or generation-oriented pipelines on RAG24. These systems exhibit larger rank displacement under persona-conditioned judging than more stable runs. The strongest instabilities are associated with smaller models such as LLaMA-3.1-8B and Qwen-2.5-7B, indicating that low-capacity evaluators amplify persona-induced ranking sensitivity.

Finally, persona source has a secondary effect compared with assessor role and judge-model capacity. Abstract PersonaHub personas and skill-grounded USPersona profiles produce broadly similar behavior, with differences concentrated in specific role--model combinations rather than forming consistent source-level effects. USPersona Orthogonal induces the largest average rank displacement on both datasets, while PersonaHub Orthogonal shows elevated displacement mainly on RAG24. This suggests that contrastive skill-grounded perspectives function as particularly strong evaluator-sensitivity probes.

Overall, the results position persona conditioning as a diagnostic mechanism for exposing assessor sensitivity rather than as a universally beneficial judging strategy. When applied to sufficiently capable models, assessor perspectives reveal where retrieval evaluation is sensitive to framing, interpretation, and evaluator assumptions while largely preserving global ranking structure. This makes persona-conditioned judging useful for stress-testing LLM-based IR evaluation pipelines and identifying evaluator-sensitive retrieval systems and architectures.
These findings suggest that controlled assessor variation can complement single-configuration LLM judging by revealing where evaluation outcomes are sensitive to assessor perspective.
\vspace{-7pt}
\section*{GenAI Usage Disclosure}
Generative AI tools were used to support manuscript editing, wording refinement, grammar checking, and LaTeX formatting. They were also used to assist with code debugging and implementation support during analysis. All research design, experimental setup, data processing decisions, analyses, results, interpretations, and scholarly claims were conducted, verified, and approved by the authors. 

\balance
\bibliographystyle{ACM-Reference-Format}
\bibliography{references}

\end{document}

%% file: win_rate_table.tex
\begin{table*}[!htbp]
\centering
\mycaption{Win--rate analysis relative to UMBRELA. W\% is the proportion of queries with higher human agreement than UMBRELA, and $\bar{\Delta}$ is the mean per-query weighted-$\kappa$ shift (with positive values indicating a shift {\emph{toward}} human preferences). Bold values indicate the highest W\% per model row.}
\label{tab:winrate_persona}
\vspace{-2mm}
\setlength{\tabcolsep}{3pt}

{\centering\textbf{TREC Deep Learning 2020}\par}
\vspace{0.8mm}
\begin{tabular}{lrrrrrr rrrrrr rrrr}
\toprule
\multirow{2}{*}{\textbf{Model}} & \multicolumn{6}{c}{\texttt{PersonaHub}} & \multicolumn{6}{c}{\texttt{USPersona}} & \multicolumn{2}{c}{\multirow{2}{*}{\textbf{Evidence}}} &  \multicolumn{2}{c}{\multirow{2}{*}{\textbf{GAP}}} \\
\cmidrule(lr){2-7}\cmidrule(lr){8-13}
& \multicolumn{2}{c}{\bf Query} & \multicolumn{2}{c}{\bf Domain} & \multicolumn{2}{c}{\bf Orthogonal} & \multicolumn{2}{c}{\bf Query} & \multicolumn{2}{c}{\bf Domain} & \multicolumn{2}{c}{\bf Orthogonal} &  &  \\
\cmidrule(lr){2-17}
& \multicolumn{1}{c}{W\%} & \multicolumn{1}{c}{$\bar{\Delta}$} & \multicolumn{1}{c}{W\%} & \multicolumn{1}{c}{$\bar{\Delta}$}& \multicolumn{1}{c}{W\%} & \multicolumn{1}{c}{$\bar{\Delta}$}& \multicolumn{1}{c}{W\%} & \multicolumn{1}{c}{$\bar{\Delta}$}& \multicolumn{1}{c}{W\%} & \multicolumn{1}{c}{$\bar{\Delta}$}& \multicolumn{1}{c}{W\%} & \multicolumn{1}{c}{$\bar{\Delta}$}& \multicolumn{1}{c}{W\%} & \multicolumn{1}{c}{$\bar{\Delta}$}& \multicolumn{1}{c}{W\%} & \multicolumn{1}{c}{$\bar{\Delta}$} \\
\midrule
GPT-4o        & 27.8 & -0.051 & \textbf{37.0} & -0.045 & 27.8 & -0.056 & 20.4 & -0.056 & 24.1 & -0.043 & 25.9 & -0.049 & 24.1 & -0.080 & 24.1 & -0.074 \\
GPT-4o-mini   & 48.1 & -0.002 & 48.1 & -0.012 & 51.9 & -0.000 & 53.7 & -0.010 & 51.9 & 0.012  & 50.0 & -0.005 & 51.9 & 0.009  & \textbf{61.1} & 0.013 \\
LLaMA3.1-70B  & 51.9 & -0.003 & 57.4 & 0.011  & 57.4 & 0.003  & 50.0 & -0.001 & \textbf{70.4} & 0.037  & 64.8 & 0.017  & 68.5 & 0.016  & 33.3 & -0.051 \\
LLaMA3.1-8B   & 7.4 & -0.123  & 7.4 & -0.136  & \textbf{9.3} & -0.123  & 3.7 & -0.153  & 1.9 & -0.163  & 7.4 & -0.140  & 3.7 & -0.102  & 0.0 & -0.148 \\
Qwen2.5-72B   & 59.3 & 0.010  & 68.5 & 0.018  & 63.0 & 0.021  & 68.5 & 0.020  & 63.0 & 0.010  & \textbf{74.1} & 0.039  & 64.8 & 0.006  & 29.6 & -0.024 \\
Qwen2.5-7B    & \textbf{63.0} & 0.009  & 61.1 & 0.019  & 40.7 & -0.087 & 53.7 & -0.044 & 42.6 & -0.051 & 31.5 & -0.193 & 59.3 & 0.023  & \textbf{63.0} & 0.015 \\
\bottomrule
\end{tabular}

\vspace{2mm}
{\centering\textbf{TREC RAG 2024}\par}
\vspace{0.5mm}
\begin{tabular}{lrrrrrr rrrrrr rrrr}
\toprule
\multirow{2}{*}{\textbf{Model}} & \multicolumn{6}{c}{\texttt{PersonaHub}} & \multicolumn{6}{c}{\texttt{USPersona}} & \multicolumn{2}{c}{\multirow{2}{*}{\textbf{Evidence}}} &  \multicolumn{2}{c}{\multirow{2}{*}{\textbf{GAP}}} \\
\cmidrule(lr){2-7}\cmidrule(lr){8-13}
& \multicolumn{2}{c}{\bf Query} & \multicolumn{2}{c}{\bf Domain} & \multicolumn{2}{c}{\bf Orthogonal} & \multicolumn{2}{c}{\bf Query} & \multicolumn{2}{c}{\bf Domain} & \multicolumn{2}{c}{\bf Orthogonal} &  &  \\
\cmidrule(lr){2-17}
& \multicolumn{1}{c}{W\%} & \multicolumn{1}{c}{$\bar{\Delta}$} & \multicolumn{1}{c}{W\%} & \multicolumn{1}{c}{$\bar{\Delta}$}& \multicolumn{1}{c}{W\%} & \multicolumn{1}{c}{$\bar{\Delta}$}& \multicolumn{1}{c}{W\%} & \multicolumn{1}{c}{$\bar{\Delta}$}& \multicolumn{1}{c}{W\%} & \multicolumn{1}{c}{$\bar{\Delta}$}& \multicolumn{1}{c}{W\%} & \multicolumn{1}{c}{$\bar{\Delta}$}& \multicolumn{1}{c}{W\%} & \multicolumn{1}{c}{$\bar{\Delta}$}& \multicolumn{1}{c}{W\%} & \multicolumn{1}{c}{$\bar{\Delta}$} \\
\midrule
GPT-4o        & 14.0 & -0.087 & \textbf{25.6} & -0.072 & 17.4 & -0.081 & 14.0 & -0.091 & 15.1 & -0.091 & 12.8 & -0.095 & 17.4 & -0.116 & 19.8 & -0.090 \\
GPT-4o-mini   & 57.0 & -0.001 & 60.5 & 0.003  & 46.5 & -0.007 & 53.5 & -0.003 & 58.1 & 0.004  & 47.7 & -0.006 & \textbf{61.6} & 0.005  & 60.5 & 0.005 \\
LLaMA3.1-70B  & 39.5 & -0.015 & 50.0 & 0.001  & 50.0 & -0.013 & 43.0 & -0.006 & 48.8 & 0.005  & 51.2 & -0.006 & \textbf{57.0} & 0.008  & 39.5 & -0.011 \\
LLaMA3.1-8B   & 27.9 & -0.059 & \textbf{30.2} & -0.048 & 27.9 & -0.056 & 20.9 & -0.071 & 23.3 & -0.073 & 23.3 & -0.061 & 27.9 & -0.047 & 20.9 & -0.079 \\
Qwen2.5-72B   & 45.3 & -0.014 & 55.8 & -0.000 & 54.7 & -0.000 & 43.0 & -0.009 & 51.2 & -0.007 & 52.3 & -0.002 & \textbf{58.1} & -0.001 & 44.2 & -0.003 \\
Qwen2.5-7B    & 38.4 & -0.013 & 45.3 & -0.008 & 31.4 & -0.077 & 37.2 & -0.053 & 38.4 & -0.045 & 12.8 & -0.173 & \textbf{47.7} & -0.014 & 40.7 & -0.021 \\
\bottomrule
\end{tabular}
\vspace{-2mm}
\end{table*}

%% file: system_stability_tau_rbo.tex
\begin{table*}[h]
\centering
\mycaption{Agreement between LLM-derived and human system rankings based on NDCG@10, measured using Kendall's $\tau$ and RBO $(\phi=0.9)$ across datasets, models, and assessor perspectives. PersonaHub and USPersona variants are grouped by Query (Q), Domain (D), and Orthogonal (O) prompts; UMB. denotes UMBRELA.}
\label{tab:system_stability_tau_rbo}
\scriptsize
\setlength{\tabcolsep}{2.2pt}
\renewcommand{\arraystretch}{0.9}
\resizebox{\textwidth}{!}{%
\begin{tabular}{ll*{18}{c}}
\toprule
\multirow{4}{*}{\textbf{Dataset}} & \multirow{4}{*}{\textbf{Metric}} 
& \multicolumn{9}{c}{\textbf{GPT-4o}} 
& \multicolumn{9}{c}{\textbf{GPT-4o-mini}} \\
\cmidrule(lr){3-11}\cmidrule(lr){12-20}
& & 
\multicolumn{3}{c}{\texttt{PersonaHub}} & \multicolumn{3}{c}{\texttt{USPersona}} & \multirow{2}{*}{\textbf{Evidence}} & \multirow{2}{*}{\textbf{GAP}} & \multirow{2}{*}{\textbf{UMB.}}
& \multicolumn{3}{c}{\texttt{PersonaHub}} & \multicolumn{3}{c}{\texttt{USPersona}} & \multirow{2}{*}{\textbf{Evidence}} & \multirow{2}{*}{\textbf{GAP}} & \multirow{2}{*}{\textbf{UMB.}} \\
\cmidrule(lr){3-5}\cmidrule(lr){6-8}\cmidrule(lr){12-14}\cmidrule(lr){15-17}
& & \bf Q & \bf D & \bf O & \bf Q & \bf D & \bf O & & & & \bf Q & \bf D & \bf O & \bf Q & \bf D & \bf O & & & \\
\midrule
\multirow{2}{*}{DL20} & $\tau$ & .917 & .924 & \textbf{.952} & .924 & .933 & .905 & .894 & .866 & .937 & .932 & .938 & .936 & \textbf{.951} & .932 & .937 & \textbf{.951} & .939 & .928 \\
     & RBO    & .866 & .702 & .707 & .697 & .780 & .626 & \textbf{.876} & .555 & .698 & .710 & .706 & .697 & .713 & .697 & .705 & \textbf{.788} & .703 & .679 \\[1ex]
\multirow{2}{*}{RAG24} & $\tau$ & .882 & .930 & \textbf{.944} & .913 & .898 & .875 & .857 & .894 & .903 & .901 & .899 & .903 & \textbf{.922} & .896 & .901 & .916 & .918 & .913 \\
      & RBO    & .604 & .644 & \textbf{.985} & .641 & .641 & .606 & .605 & .642 & .665 & \textbf{.665} & .642 & .643 & .641 & .640 & .664 & .642 & .641 & .641 \\

\midrule
\multirow{4}{*}{\textbf{Dataset}} & \multirow{4}{*}{\textbf{Metric}} 
& \multicolumn{9}{c}{\textbf{LLaMA-3.1-70B}} 
& \multicolumn{9}{c}{\textbf{LLaMA-3.1-8B}} \\
\cmidrule(lr){3-11}\cmidrule(lr){12-20}
& & 
\multicolumn{3}{c}{\texttt{PersonaHub}} & \multicolumn{3}{c}{\texttt{USPersona}} & \multirow{2}{*}{\textbf{Evidence}} & \multirow{2}{*}{\textbf{GAP}} & \multirow{2}{*}{\textbf{UMB.}}
& \multicolumn{3}{c}{\texttt{PersonaHub}} & \multicolumn{3}{c}{\texttt{USPersona}} & \multirow{2}{*}{\textbf{Evidence}} & \multirow{2}{*}{\textbf{GAP}} & \multirow{2}{*}{\textbf{UMB.}} \\
\cmidrule(lr){3-5}\cmidrule(lr){6-8}\cmidrule(lr){12-14}\cmidrule(lr){15-17}
& & \bf Q & \bf D & \bf O & \bf Q & \bf D & \bf O & & & & \bf Q & \bf D & \bf O & \bf Q & \bf D & \bf O & & & \\
\midrule
\multirow{2}{*}{DL20} & $\tau$ & .935 & .939 & .938 & \textbf{.946} & .944 & .911 & .938 & .936 & .932 & .707 & .814 & .849 & .741 & .815 & .716 & .627 & .721 & \textbf{.902} \\
     & RBO    & .715 & .736 & .736 & .707 & \textbf{.742} & .699 & .716 & .735 & .708 & .393 & .652 & \textbf{.655} & .613 & .457 & .348 & .542 & .393 & .648 \\[1ex]
\multirow{2}{*}{RAG24} & $\tau$ & .944 & .946 & .922 & \textbf{.958} & .944 & .951 & .951 & .927 & .900 & .815 & .812 & .749 & .792 & .784 & .741 & .824 & .807 & \textbf{.853} \\
      & RBO    & .991 & .992 & .644 & .994 & .983 & \textbf{.995} & .992 & .991 & .639 & .987 & .988 & .290 & .588 & .803 & .288 & \textbf{.989} & .883 & .803 \\

\midrule
\multirow{4}{*}{\textbf{Dataset}} & \multirow{4}{*}{\textbf{Metric}} 
& \multicolumn{9}{c}{\textbf{Qwen-2.5-72B}} 
& \multicolumn{9}{c}{\textbf{Qwen-2.5-7B}} \\
\cmidrule(lr){3-11}\cmidrule(lr){12-20}
& & 
\multicolumn{3}{c}{\texttt{PersonaHub}} & \multicolumn{3}{c}{\texttt{USPersona}} & \multirow{2}{*}{\textbf{Evidence}} & \multirow{2}{*}{\textbf{GAP}} & \multirow{2}{*}{\textbf{UMB.}}
& \multicolumn{3}{c}{\texttt{PersonaHub}} & \multicolumn{3}{c}{\texttt{USPersona}} & \multirow{2}{*}{\textbf{Evidence}} & \multirow{2}{*}{\textbf{GAP}} & \multirow{2}{*}{\textbf{UMB.}} \\
\cmidrule(lr){3-5}\cmidrule(lr){6-8}\cmidrule(lr){12-14}\cmidrule(lr){15-17}
& & \bf Q & \bf D & \bf O & \bf Q & \bf D & \bf O & & & & \bf Q & \bf D & \bf O & \bf Q & \bf D & \bf O & & & \\
\midrule
\multirow{2}{*}{DL20} & $\tau$ & .926 & \textbf{.942} & .936 & .929 & .935 & .921 & .939 & .938 & .936 & .890 & .903 & .868 & .869 & \textbf{.932} & .812 & .892 & .897 & .896 \\
     & RBO    & .700 & .699 & .712 & .698 & .723 & .645 & \textbf{.769} & .706 & .693 & .633 & .628 & \textbf{.852} & .550 & .762 & .617 & .771 & .713 & .746 \\[1ex]
\multirow{2}{*}{RAG24} & $\tau$ & .908 & .900 & .892 & .944 & .910 & .908 & \textbf{.953} & .901 & .902 & .930 & \textbf{.968} & .867 & .960 & .898 & .834 & .914 & .919 & .925 \\
      & RBO    & .663 & .640 & .638 & .991 & .663 & .663 & \textbf{.992} & .639 & .640 & .643 & \textbf{.892} & .294 & .890 & .641 & .586 & .640 & .641 & .662 \\
\bottomrule
\end{tabular}%
}
\end{table*}